\documentclass[11pt,epsf]{article}
\usepackage{amsmath}
\usepackage{amsfonts}
\usepackage{amssymb}
\usepackage{graphicx}
\usepackage{color}
\usepackage{comment}

\newtheorem{theorem}{Theorem}
\newtheorem{lemma}{Lemma}
\newcommand {\dfn} {\stackrel{\Delta} {=}}
\newcommand {\exe} {\stackrel{\cdot} {=}}
\newcommand {\lexe} {\stackrel{\cdot} {\le}}

\newcommand{\eqa}{\stackrel{\mbox{\tiny (a)}}{=}}

\newcommand {\reals} {{\rm I\!R}}

\newcommand {\bx} {\mbox{\boldmath $x$}}

\newcommand {\bE} {\mbox{\boldmath $E$}}

\newcommand {\bX} {\mbox{\boldmath $X$}}

\newcommand{\calC}{{\cal C}}
\newcommand{\calD}{{\cal D}}
\newcommand{\calE}{{\cal E}}

\newcommand{\calI}{{\cal I}}

\newcommand{\calQ}{{\cal Q}}

\newcommand{\calX}{{\cal X}}

\newcommand{\hx}{{\hat{x}}}
\newcommand {\hbx} {\mbox{\boldmath $\hat{x}$}}
\allowdisplaybreaks

\begin{document}
\thispagestyle{empty}
\title{$D$-Semifaithful Codes that are Universal over Both Memoryless Sources and
Distortion Measures}
\author{Neri Merhav}
\date{}
\maketitle

\begin{center}
The Andrew \& Erna Viterbi Faculty of Electrical and Computer Engineering\\
Technion - Israel Institute of Technology \\
Technion City, Haifa 32000, ISRAEL \\
E--mail: {\tt merhav@ee.technion.ac.il}\\
\end{center}
\vspace{1.5\baselineskip}
\setlength{\baselineskip}{1.5\baselineskip}

\begin{abstract}
We prove the existence of codebooks for $d$-semifaithful lossy compression that are
simultaneously universal with respect to both the class of finite-alphabet memoryless sources 
and the class of all bounded additive distortion measures. By applying
independent random selection of the codewords according to a mixture of all
memoryless sources, we achieve redundancy rates that are within $O(\log n/n)$
close to the empirical rate-distortion function of every given source vector
with respect to every bounded distortion measure.
As outlined in the last section, the principal ideas can also be extended significantly beyond the class of memoryless
sources, namely, to the setting of individual sequences encoded by finite-state
machines.\\

\noindent
{\bf Index Terms:} lossy compression, rate-distortion theory, universal
coding, random coding, Lempel-Ziv algorithm.
\end{abstract}

\clearpage
\section{Introduction}

We consider the classical problem of lossy compression for finite-alphabet
memoryless sources with respect to a fidelity
criterion defined by an additive distortion measure \cite{Berger71}, \cite[Chap.\ 10]{CT06}, \cite[Chap.\
9]{Gallager68}, \cite{Gray90}, \cite[Chaps.\ 7,8]{VO79}. More specifically,
our focus is on $d$-semifaithful codes, i.e., variable--length codes that meet a given
distortion constraint for each and every source sequence (and not only on the
average). As is very well known \cite{Berger71}, the rate-distortion function
characterizes the least achievable expected coding rate for a given
memoryless source and distortion measure.

Motivated by the consideration that the source statistics are seldom
known in practice, many research efforts, throughout the years, have been
devoted to the quest for
universal codes, namely, codes that are independent of the unknown memoryless
source, but nevertheless, achieve the rate-distortion function asymptotically,
for long blocks, see, e.g., \cite{Kontoyiannis00}, \cite{MW22a}, \cite{MW22b},
\cite{me93}, \cite{OS90}, \cite{SP21}, \cite{YZ01}, \cite{YS93}, which is by no means an
exhaustive list of
all relevant articles. This line of research, along with its various types of
universality (weak universality, strong universality, expected vs.\ almost-sure
convergence, etc.) complements and partially extends its lossless
counterpart, yet it should be pointed out that the theory of universal
lossless source coding is
significantly more mature and well developed, along with ties to other
problem areas, such as channel capacity theory and universal prediction theory
(see, for example, \cite{MF98}).

In a recent work coauthored with Cohen \cite{CM21} (which is a further
development over \cite{AM98} and \cite{MC20}), we considered the intimately
related problem of
universal guessing subject to a fidelity criterion, where the universality
takes place in a multitude of dimensions. One of those dimensions is the distortion
measure. In this paper, the ideas
of \cite{CM21} are harnessed and considerably refined to demonstrate the existence of
$d$-semifaithful codes,
which are not only universal with respect to (w.r.t.) the 
source statistics, but also universal w.r.t.\ the class of all bounded single--letter
distortion measures. In other words, the same universal codebook is completely
flexible to be used, not only for
one given distortion measure, but for all bounded distortion measures,
on the top of its universality property for all memoryless sources of a given alphabet, as before.
This means that it is enough that the distortion measure would be specified
once a source vector has to be actually encoded, and not necessarily before the codebook is
constructed. Recently, Mahmood and Wagner have also provided very
interesting results along the very same line \cite{MW22a},
\cite{MW22b}. In \cite{MW22a}, they proposed three universal coding schemes.
The first two are based on unions of codebooks associated with distortion measures
that belong to a fine grid in the space of all bounded
distortion matrices. The third scheme is based on the notion of the
Vapnik-Chervonenkis (VC) dimension \cite{Vapnik98}. All three coding schemes
achieve rate redundancies that are asymptotically proportional to $\frac{\log
n}{n}$ for blocks of
length $n$, but they differ in the constants of proportionality. In
\cite{MW22b}, as its title suggests, the focus is more towards strong universality 
and minimax properties of universal codes. Accordingly, several coding
theorems are provided in \cite{MW22b}, but the uniformity comes at the
inevitable price
of a slowdown in the decay of the rate redundancies. 

Our approach
is conceptually much simpler than those of \cite{MW22a} and \cite{MW22b}, and we
show that smaller rate redundancies are achievable. Moreover, the analysis is also
simpler, as its main part is based on a saddle-point derivation of the
probability that a randomly selected codeword would fall within distortion $nD$
away from a source sequence of a given type class. This bound is
asymptotically tight in the sense that, it does not only have the correct
exponential behavior, but moreover, the ratio between the bound and the
exact probability tends to unity as the block length $n$ grows without bound.
However, for the sake of fairness, it must be pointed out that in contrast to
\cite{MW22b}, we make no
claims concerning uniformity of convergence. Finally, we provide an informal
outline of an extension of the main
ideas beyond the realm
of memoryless sources and additive distortion measures, as we consider individual source sequences
encoded by finite-state machines, in the spirit of the Lempel-Ziv setting
\cite{ZL78}.

The outline of the remaining part of this paper is as follows. In Section
\ref{nps}, we establish the notation and formalize the problem.
In Section \ref{mainlemma}, we state and prove a lemma that provides an
asymptotically tight evaluation of
the probability that a random codeword happens to lie within
distortion $nD$ away from the source vector. In Section \ref{mainresult}, we state
and prove the main coding theorem concerning the universality. Finally, in
Section \ref{LZ}, we consider the broader setup mentioned above.

\section{Notation and Problem Setting}
\label{nps}

Throughout the paper, random variables will be denoted by capital
letters, specific values they may take will be denoted by the
corresponding lower case letters, and their alphabets
will be denoted by calligraphic letters. Random
vectors and their realizations will be denoted,
respectively, by capital letters and the corresponding lower case letters,
both in the bold face font. Their alphabets will be superscripted by their
dimensions. For example, the random vector $\bX=(X_1,\ldots,X_n)$, ($n$ --
positive integer) may take a specific vector value $\bx=(x_1,\ldots,x_n)$
in $\calX^n$, the $n$--th order Cartesian power of $\calX$, which is
the alphabet of each component of this vector.
Sources and channels will be denoted by the letter $P$ or $Q$.
The probability of an event $\calE$ will be denoted by $\mbox{Pr}\{\calE\}$,
and the expectation
operator with respect to (w.r.t.) a probability distribution $P$ will be
denoted by
$\bE\{\cdot\}$. 
For two positive sequences, $a_n$ and $b_n$, the notation $a_n\exe b_n$ will
stand for equality in the exponential scale, that is,
$\lim_{n\to\infty}\frac{1}{n}\log \frac{a_n}{b_n}=0$. 
Similarly,
$a_n\lexe b_n$ means that
$\limsup_{n\to\infty}\frac{1}{n}\log \frac{a_n}{b_n}\le 0$, and so on.
The notation $a_n\sim b_n$, for two positive sequences, will stand for the property that 
$\lim_{n\to\infty}\frac{a_n}{b_n}=1$.
The indicator function
of an event $\calE$ will be denoted by $\calI\{E\}$. The notation $[x]_+$
will stand for $\max\{0,x\}$.
The logarithmic function, $\log x$, will be understood to be defined to the
base 2. Logarithms to the base $e$ will be denote by $\ln$.
The empirical distribution of a sequence $\bx\in\calX^n$, which will be
denoted by $\hat{P}_{\bx}$, is the vector of relative frequencies
$\hat{P}_{\bx}(x)$
of each symbol $x\in\calX$ in $\bx$.

Let $X_1,X_2,\ldots$ be independent copies of a random variable
(RV) $X$, taking on values in a finite alphabet $\calX=\{1,2,\ldots,K\}$,
where $K > 1$ is a positive integer. We denote the distribution of $X$ by
$P=\{P(x),~x\in\calX\}$, where $P(x)\dfn \mbox{Pr}\{X=x\}$. Let
$\hat{\calX}=\{1,2,\ldots,J\}$ denote a finite reconstruction alphabet, where
$J > 1$ is also a positive integer. A distortion measure
$d:\calX\times\hat{\calX}\to\reals^+$ is a non-negative function of pairs
$(x,\hx)\in\calX\times\hat{\calX}$, which can also be thought of as a $K\times
J$ matrix whose $(j,k)$-th entry is given by $d(j,k)$, $1\le j\le J$, $1\le
k\le K$. We assume that the distortion measure $d$ satisfies two requirements:
\begin{itemize}
\item [(i)] For every $x\in\calX$, $\min_{\hx}d(x,\hx)=0$; 
\item [(ii)]
$d_{\max}\dfn\max_{(x,\hx)\calX\times\hat{\calX}}d(x,\hx)<\infty$. 
\end{itemize}
Note that (i) does not limit the
generality, as every distortion measure can be modified so as to satisfy (i)
without changing the essence. This is done
by defining $d'(x,\hx)=d(x,\hx)-\min_{\hx}d(x,\hx)$, shifting the
distortion level $D$ to $D'=D-\bE\{\min_{\hx}d(X,\hx)\}$, and observing that
the shift, $\bE\{\min_{\hx}d(X,\hx)\}$, depends only on the source $P$ and the
distortion measure, not on the code.
The distortion between two vectors $\bx\in\calX^n$ and $\hbx\in\hat{\calX}^n$
will be defined additively as
\begin{equation}
d(\bx,\hbx)=\sum_{i=1}^nd(x_i,\hx_i).
\end{equation}

A block code of length $n$ consists of an encoder and a
decoder. We consider a variable-rate encoder, which is a mapping,
$\phi_n:\calX^n\to\{0,1\}^*$, that maps the space of source vectors of length
$n$, $\calX^n$, into a set, $\{0,1\}^*$, of variable-length compressed bit
strings. The decoder is a mapping, $\psi_n:\{0,1\}^*\to \calC_n\subseteq\hat{\calX}^n$,
that maps the space of compressed strings into a codebook, $\calC_n$, which is
a certain subset of the reproduction space, $\hat{\calX}^n$. The length (in
nats) of $\phi_n(\bx)$ will be denoted by $L_d(\bx)$, where the subscript $d$
denotes the distortion measure.\footnote{The need for the subscript $d$ will
become clear in the sequel.} The coding rate for $\bx$ is
$L_d(\bx)/n$.

A code is called
$d$-semifaithful w.r.t.\ a given distortion level $D$, if for every
$\bx\in\calX^n$, 
\begin{equation}
d(\bx,\psi_n(\phi_n(\bx)))\le nD. 
\end{equation}
As is well known, the rate-distortion coding
theorem asserts that for a given memoryless source $P$ and distortion measure
$d$, there exist $d$-semifaithful codes, $(\phi_n,\psi_n)$, w.r.t.\ distortion level $D$,
whose average coding rate $R$ is arbitrarily close to
\begin{equation}
R_d(D,P)\dfn\min_{\{P_{\hat{X}|X}:~\bE\{d(X,\hat{X})\le D\}}I(X;\hat{X}),
\end{equation}
for all sufficiently large $n$.
On the other hand, the converse theorem asserts that there are no $d$-semifaithful
codes w.r.t.\ distortion level $D$ with $R < R_d(D,P)$.

The following Lagrange-dual representation of $R_d(D,P)$ (in nats per source
symbol) is well
known (see, e.g., \cite[p.\ 90, Corollary 4.2.3]{Gray90}):
\begin{eqnarray}
R_d(D,P)&=&\sup_{s\ge
0}\min_Q\left\{-\sum_{x\in\calX}P(x)\ln\left[\sum_{\hx\in\hat{\calX}}Q(\hx)e^{-sd(x,\hx)}\right]-sD\right\}\nonumber\\
&=&\min_Q\sup_{s\ge 0}\left\{-\sum_{x\in\calX}P(x)\ln\left[\sum_{\hx\in\hat{\calX}}Q(\hx)e^{-sd(x,\hx)}\right]-sD\right\},
\end{eqnarray}
where minimization is over all probability assignments,
$Q=\{Q(\hx),~\hx\in\hat{\calX}\}$, across the reproduction alphabet,
$\hat{\calX}$. Here, 
the second equality holds since the function,
\begin{equation}
\label{FsQ}
F(s,Q)\dfn-\sum_{x\in\calX}P(x)\ln\left[\sum_{\hx\in\hat{\calX}}Q(\hx)e^{-sd(x,\hx)}\right]-sD
\end{equation}
is convex in $Q$ and concave in $s$.

Our objective is to prove that there exists a sequence of codes,
$\{(\phi_n,\psi_n)\}_{n\ge 1}$ that are simultaneously
$d$-semifaithful w.r.t.\ $D$ for every distortion measure $d$ that satisfies
requirements (i) and (ii), and, at the same time, their code-length functions are arbitrarily close to
$R_d(D,\hat{P}_{\bx})$ for all $\bx\in\calX^n$ when $n$ is sufficiently large $n$. We will also focus on the achievable redundancy as a
function of $n$. 

\section{The Probability of a Successful Single Random Selection}
\label{mainlemma}

This section is devoted to a lemma that stands at the heart of the derivations
in this work: It provides an asymptotically tight assessment of the
probability that a single randomly selected codeword happens to fall within
distortion no more than $nD$ away from a given source vector $\bx\in\calX^n$,
which has a certain empirical distribution, $\hat{P}_{\bx}$. The concept of
proving achievability of $R_d(D,P)$ via the such a lower bound is, of course, by no means new,
and it serves as the classical tool for proving the direct part of the
rate-distortion coding theorem. 
There are two points, however, that make our derivation somewhat different
from the traditional one.
\begin{enumerate}
\item We select a universal random coding distribution that is asymptotically
as good as the optimal one for every source and every distortion measure.
\item Our analysis is based upon the saddle-point method
(a.k.a.\ the steepest descent method) \cite[Chap.\
5]{debruijn81}, \cite[Section 4.3]{itsp09}, which is not
only exponentially tight, but moreover, it is asymptotically tight in the sense that the
ratio between the approximate expression and the exact probability tends to unity as
$n\to\infty$. As a consequence, it gives rise to a precise characterization of
the redundancy terms as well.
\end{enumerate}

Consider the random coding distribution, given by the uniform\footnote{The
choice of the uniform mixture is motivated merely by its convenience. It can
be replaced by any density $w(Q)$, as long as it is bounded away from
zero and from infinity.} mixture of all
memoryless sources,
\begin{equation}
\label{mixture}
W(\hbx)=(J-1)!\cdot\int_{\calQ}\mbox{d}Q\cdot\prod_{i=1}^nQ(\hx_i),
\end{equation}
where $\calQ$ is the simplex of all probability assignments over $\hat{\calX}$
and the factor $(J-1)!$ is a normalization constant that accounts
for the fact the volume of $\calQ$
is $1/(J-1)!$.\footnote{This well known fact can easily be proved either by induction
on $J$ or by the simple observation that the volume occupied by the set of vectors,
$(u_1,\ldots,u_{J-1})$, with ordered components, $0\le u_1\le u_2\le\ldots\le
u_{J-1}\le 1$, which is obviously $1/(J-1)!$, can be transformed bijectively
into a set of $J-1$ probabilities, $p_1=u_1$, $p_2\le u_2-u_1$, ...,
$p_{J-1}=u_{J-1}-u_{J-2}$ (whose sum is $u_{J-1}\le 1$), and that the Jacobian
of this transformation is $1$, so it does not alter the volume.}
The probability of a {\em successful single random selection}, for a given
source sequence $\bx$, is defined as
\begin{equation}
P_{\mbox{\tiny
s}}^d[\bx]\dfn\sum_{\{\hbx:~d(\bx,\hbx)\le
nD\}}W(\hbx)=(J-1)!\cdot\int_{\calQ}\mbox{d}Q\sum_{\{\hbx:~d(\bx,\hbx)\le nD\}}\prod_{i=1}^nQ(\hx_i).
\end{equation}
Before stating our main lemma, we need a few more definitions.\\
\noindent
1. For the case where the non-zero entries of the distortion matrix,
$\{d(j,k),~1\le j\le J,~1\le k\le K\}$,
are all commensurable, i.e., the ratios, $d(j,k)/d(j',k')$ ($(j',k')\ne (j,k)$,
$d(j',k')> 0$) are all rational numbers,
we define $\Delta$ as the greatest common factor of $\{d(j,k):~d(j,k)> 0,~1\le j\le J,~1\le
k\le K\}$. In other words, $\Delta$ is the largest positive real, $\delta$, such that
$d(j,k)/\delta$ is a positive integer for every $(j,k)$ with $d(j,k)> 0$.
Otherwise, if the non-zero entries of the distortion matrix are
incommensurable, we define $\Delta=0$ (which amounts to passing to the limit
$\Delta\to 0$).\\
\noindent
2. For a given $Q\in\calQ$, let
$s_0$ be the (unique) maximizer of $F(s,Q)$ (defined in eq.\ (\ref{FsQ})) in the range $s\ge 0$, which is
given as follows. If $D < D_{\max}(Q)\dfn\sum_{x,\hx}P(x)Q(\hx)d(x,\hx)$, then
$s_0$ is the solution $s$ to the equation
\begin{equation}
\sum_xP(x)\cdot\frac{\sum_{\hx}Q(\hx)e^{-s
d(x,\hx)}d(x,\hx)}{\sum_{\hx}Q(\hx)e^{-sd(x,\hx)}}=D.
\end{equation}
Note that $s_0$ depends on $Q$, and accordingly, in the sequel, we will denote it
sometimes as $s_0(Q)$, especially in places where it will be important to emphasize this dependence.
If $D\ge D_{\max}(Q)$, $s_0=0$.
For $s> 0$, we define $M(s,Q)$ as the absolute value of the second derivative of $F(s,Q)$
w.r.t.\ $s$.
Let $Q_0$
be the minimizer of $F(s_0(Q),Q)$. For
$D < D_{\max}(Q_0)$, we define $|\mbox{Hess}_F(Q_0)|$ as the determinant of
the $(J-1)\times(J-1)$ Hessian matrix of $F(s_0(Q),Q)$ w.r.t.\ the (first) $J-1$
components of $Q$, computed at $Q=Q_0$.
Finally, define the function
\begin{equation}
\label{K}
K_n[s,Q]\dfn
\frac{\Delta\exp\{-s[(nD)~\mbox{mod}~\Delta]\}}{(1-e^{-s\Delta})\sqrt{2\pi M(s,Q)}},
\end{equation}
where $a~\mbox{mod}~b\dfn a-b\cdot\lfloor a/b\rfloor$.
We are now ready to state the following lemma.
\begin{lemma}
\label{lemma1}
Let the assumptions of Section \ref{nps} hold. Then,
\begin{equation}
P_{\mbox{\tiny s}}^d[\bx]\sim\left\{\begin{array}{ll}
(J-1)!\cdot\frac{(2\pi)^{(J-1)/2}K_n[s_0(Q_0),Q_0]}{\sqrt{|\mbox{Hess}_F(Q_0)|}}\cdot\frac{\exp\{-nR_d(D,\hat{P}_{\bx})\}}{n^{J/2}}
& R_d(D,\hat{P}_{\bx})> 0\\
(J-1)!\cdot\mbox{Vol}\{Q:~D_{\max}(Q)\le D\}\cdot[1-o(n)] &
R_d(D,\hat{P}_{\bx})=0,\end{array}\right.
\end{equation}
where $\hat{P}_{\bx}$ denotes the empirical distribution of $\bx\in\calX^n$.
\end{lemma}

\noindent
{\em Discussion.}
A few comments are in order concerning Lemma \ref{lemma1}.

\noindent
1. First, a technical issue should be clarified.
Note that although the factor $K_n[s_0(Q_0),Q_0]$ depends on $n$, it does not
tend to zero as $n\to\infty$ and hence does not affect the asymptotic behavior
for large $n$. Referring to eq.\ (\ref{K}), this is easily seen by observing 
that the only dependence on $n$ is in the exponential term of the numerator, which oscillates between $e^{-s\Delta}$ and $1$.
We therefore conclude that in the interesting case where
$R_d(D,\hat{P}_{\bx})>0$,
\begin{equation}
\label{maincorollary}
P_{\mbox{\tiny
s}}^d[\bx]\sim\exp\left\{-n\left[R_d(D,\hat{P}_{\bx})+\frac{J\ln n}{2n}
+o\left(\frac{\ln n}{n}\right)\right]\right\}.
\end{equation}
For $R_d(D,\hat{P}_{\bx})=0$,
$P_{\mbox{\tiny s}}^d[\bx]$ is essentially a positive constant.

\noindent
2. The choice of the mixture distribution (\ref{mixture}) as our random coding
distribution is inspired by earlier works on the intimately related problem of
guessing, \cite{CM21}, \cite{MC20}, but here our analysis is more refined 
for the quest of quantifying rate redundancies. For a rough insight
on the rationale behind this choice, consider the following line of thought.
Intuitively, $W(\bx)$ is exponentially
equivalent to the normalized maximum-likelihood (NML) distribution, that is proportional to
$\max_{Q\in\calQ}Q(\hbx)$, 
whose normalization factor, $\sum_{\bx}\max_{Q\in\calQ}Q(\hbx)$ (a.k.a.\ the Shtarkov sum),
grows only polynomially with $n$ (as can easily be seen by the method of types). Consequently, the probability of any $\bx$ under
the NML distribution (and hence also under $W$),
is exponentially no smaller than
$Q(\hbx)$ for every product
distribution $Q$, including the optimal one. 
As a result, the probability
of a single success under $W$ is exponentially no worse than the one induced
by every product distribution $Q$. Indeed, we could have chosen our random
distribution to be the NML distribution, but the mixture distribution, $W$, lends
itself more conveniently to analysis. In fact, Mahmood and Wagner \cite{MW22b}
employed the NML distribution, but in a different way than here.

The remaining part of this section is devoted to the proof of Lemma
\ref{lemma1}.\\

\noindent
{\em Proof of Lemma \ref{lemma1}.}
We begin with an evaluation of the probability of a single success under a
given memoryless $Q$, leaving the integration over $\calQ$ for the next step.
Our proof is based on the following identity regarding
the unit step function, $u(x)\dfn\calI\{x\ge 0\}$, which manifests the fact
that it can be represented as the inverse Laplace transform (Mellin's inverse formula) of the complex function
$1/z=\int_0^\infty e^{-zx}\mbox{d}z$ ($\mbox{Re}\{z\}>0$):
\begin{equation}
u(x)=\frac{1}{2\pi
j}\lim_{A\to\infty}\int_{c-jA}^{c+jA}\frac{e^{zx}}{z}\cdot\mbox{d}z,
\end{equation}
where $j\dfn\sqrt{-1}$ and $c$ is an arbitrary positive real.
We then have the following chain of equalities:
\begin{eqnarray}
& &\sum_{\{\hbx:~d(\bx,\hbx)\le nD\}}Q(\hbx)\nonumber\\
&=&\sum_{\hbx\in\hat{\calX}^n}Q(\hbx)\cdot
u\left(nD-\sum_{i=1}^nd(x_i,\hx_i)\right)\nonumber\\
&=&\sum_{\hbx\in\hat{\calX}^n}Q(\hbx)\cdot\frac{1}{2\pi
j}\lim_{A\to\infty}\int_{c-jA}^{c+jA}\frac{\mbox{d}z}{z}\exp\left\{z\left(nD-\sum_{i=1}^nd(x_i,\hx_i)\right)\right\}\nonumber\\
&=&\frac{1}{2\pi
j}\lim_{A\to\infty}\int_{c-jA}^{c+jA}
\frac{\mbox{d}z}{z}e^{znD}\sum_{\hbx\in\hat{\calX}^n}Q(\hbx)\cdot\exp\left\{-z\sum_{i=1}^nd(x_i,\hx_i)\right\}\nonumber\\
&=&\frac{1}{2\pi
j}\lim_{A\to\infty}\int_{c-jA}^{c+jA}\frac{\mbox{d}z}{z}e^{znD}\prod_{x\in\calX}\left[\sum_{\hx\in\calX}Q(\hx)
e^{-zd(x,\hx)}\right]^{n\hat{P}_{\bx}(x)}\nonumber\\
&=&\frac{1}{2\pi
j}\lim_{A\to\infty}\int_{c-jA}^{c+jA}\frac{\mbox{d}z}{z}\exp\left\{n\left(zD+\sum_{x\in\calX}\hat{P}_{\bx}(x)\ln\left[\sum_{\hx}Q(\hx)
e^{-zd(x,\hx)}\right]\right)\right\}\nonumber\\
&=&\frac{1}{2\pi
j}\lim_{A\to\infty}\int_{c-jA}^{c+jA}\frac{e^{-nF(z,Q)}}{z}\cdot\mbox{d}z.
\end{eqnarray}
The right--most side of this chain of equalities is an integral
of an exponential function with a large parameter $n$, along the vertical line
in the complex plane, $\mbox{Re}\{z\}=c$.
This integral will now be assessed using the saddle-point
method. 

Consider the case where $Q$ is such that $D < D_{\max}(Q)$, so that $s_0> 0$. 
Suppose first that the positive entries of the distortion matrix
are commensurable with a greatest common factor given by $\Delta> 0$.
Since all non-zero $\{d(j,k)\}$ are integer multiples of $\Delta$, the function
$|e^{-nF(s_0+j\omega,Q)}|=\exp[-n\mbox{Re}\{F(s_0+j\omega,Q)\}]$ is periodic 
in $\omega$ with period $\Omega\dfn 2\pi/\Delta$. Therefore,
in the limit of $A\to\infty$,
there are infinitely many dominant saddle-points, all of the form $z=s_0+j\Omega\ell$,
$\ell=0,\pm 1,\pm 2,\ldots$, as in all these points, $|e^{-nF(z,Q)}|$ has a local
maximum in the vertical direction of the complex plane
(which is a global maximum within each period), and a minimum along the
horizontal axis.
In order that the integration path,
$\mbox{Re}\{z\}=c$, would pass via
all saddle-points, we select $c=s_0$.
Thus, according to the saddle-point method \cite[Chap.\ 5]{debruijn81},
\cite[Sect.\ 4.3]{itsp09},
in this case, we have
\begin{eqnarray}
\label{14}
\sum_{\{\hbx:~d(\bx,\hbx)\le nD\}}Q(\hbx)
&\sim&\frac{e^{j\pi/2}}{2\pi j}\sum_{\ell=-\infty}^{\infty}\frac{\exp\{-nF(s_0+j\Omega\ell,Q)\}}{s_0+j\Omega\ell}
\cdot\sqrt{\frac{2\pi}{M(s_0+j\Omega\ell,Q)n}}\nonumber\\
&=&\left(\frac{1}{2\pi}\sum_{\ell=-\infty}^{\infty}\frac{e^{j\Omega\ell nD}}{s_0+j\Omega\ell}\right)
\cdot\exp\{-nF(s_0,Q)\}
\cdot\sqrt{\frac{2\pi}{M(s_0,Q)n}},
\end{eqnarray}
where in the asymptotic equality step, we have collected the
contributions of all dominant saddle-points along the integration path from $s_0-j
\infty$ to $s_0+j\infty$ (where the factor $e^{j\pi/2}=j$ accounts for the
vertical axis of all saddle-points), and then, in the next equality, we have used the periodicity of
$e^{-n\mbox{Re}\{F(z,Q)\}}$ (and hence also of its second derivative) in the vertical direction. 
We next address the infinite summation in the brackets of the last line of
(\ref{14}).
\begin{eqnarray}
\label{infinitesum}
\frac{1}{2\pi}\sum_{\ell=-\infty}^{\infty}\frac{e^{j\Omega\ell nD}}{s_0+j\Omega\ell}
&=&\frac{1}{2\pi}\int_{-\infty}^\infty e^{j\omega nD}\cdot\frac{1}{s_0+j\omega}\cdot\left[
\sum_{\ell=-\infty}^{\infty}\delta(\omega-\Omega\ell)\right]\mbox{d}\omega\nonumber\\
&\eqa&\left\{\left[e^{-s_0t}u(t)\right]\star\left[\frac{1}{\Omega}\sum_{k=-\infty}^\infty\delta\left(t-\frac{2\pi k}{\Omega}\right)
\right]\right\}\bigg|_{t=nD}\nonumber\\
&=&\frac{1}{\Omega}\sum_{k=-\infty}^\infty e^{-s_0(nD-2\pi k/\Omega)}u\left(nD-\frac{2\pi k}{\Omega}\right)\nonumber\\
&=&\frac{1}{\Omega}\exp\left\{-s_0\left[(nD)\mbox{mod}\left(\frac{2\pi}{\Omega}\right)\right]\right\}\cdot
\sum_{k=0}^\infty e^{-s_02\pi k/\Omega}\nonumber\\
&=&\frac{\exp\left\{-s_0\left[(nD)\mbox{mod}\left(\frac{2\pi}{\Omega}\right)\right]\right\}}{\Omega(1-e^{-2\pi s_0/\Omega})}\nonumber\\
	&=&\frac{\Delta\exp\{-s_0[(nD)~\mbox{mod}~\Delta]\}}{2\pi(1-e^{-s_0\Delta})},
\end{eqnarray}
where in (a) we have used the fact that inverse Fourier transform of the product of two frequency-domain functions 
is equal to the convolution between the
individual inverse Fourier transforms. 
If the positive distortions, $\{d(j,k)\}$, are incommensurable, then
$\mbox{Re}\{F(s_0+j\omega,Q)\}$ is no longer periodic and then only $z=s_0$
is a dominant saddle-point.
This can be viewed as a special case
pertaining to the limit $\Delta\to 0$ (or, equivalently, $\Omega\to\infty$), which matches the above formal definitions of
$\Delta$ and $\Omega$ in the incommensurable case.
On substituting the right-most side of (\ref{infinitesum}) back into (\ref{14}), we obtain
\begin{eqnarray}
\sum_{\{\hbx:~d(\bx,\hbx)\le nD\}}Q(\hbx)&\sim&
\frac{\Delta\exp\{-s_0[(nD)~\mbox{mod}~\Delta]\}}{2\pi(1-e^{-s_0\Delta})}\cdot
\sqrt{\frac{2\pi}{M(s_0,Q)n}}\cdot\exp\{-nF(s_0,Q)\}\nonumber\\
&=&K_n[s_0,Q]\cdot\frac{\exp\{-nF(s_0,Q)\}}{\sqrt{n}}.
\end{eqnarray}
In the case where $Q$ is such that $D> D_{\max}(Q)$, 
\begin{equation}
	\sum_{\{\hbx:~d(\bx,\hbx)\le nD\}}Q(\hbx)=1-\sum_{\{\hbx:~d(\bx,\hbx)> nD\}}Q(\hbx)=1-o(n),
\end{equation}
by the weak law of large numbers. However, unless $R_d(D,\hat{P}_{\bx})=0$, there is no $Q$ for which
$D> D_{\max}(Q)$. 

We now move on to the second step, of integration over $\calQ$, which will be
carried out using the multivariate version of the Laplace method of integration (see, e.g., \cite[Chap.\
4]{debruijn81}, \cite[Section 4.2]{itsp09}).
Assuming that $R_d(D,\hat{P}_{\bx})>0$,
\begin{eqnarray}
& &\sum_{\{\hbx:~d(\bx,\hbx)\le nD\}}W(\hbx)\nonumber\\
&=&(J-1)!\cdot\int_{\calQ}\sum_{\{\hbx:~d(\bx,\hbx)\le nD\}}Q(\hbx)\mbox{d}Q\nonumber\\
&\sim&(J-1)!\cdot\int_{\calQ}\mbox{d}Q\cdot\frac{K_n[s_0(Q),Q]}{\sqrt{n}}\cdot
e^{-nF(s_0(Q),Q)}\nonumber\\
&=&(J-1)!\cdot\left(\frac{2\pi}{n}\right)^{(J-1)/2}\cdot\frac{1}{\sqrt{|\mbox{Hess}_F(Q_0)}|}
\frac{K_n[s_0(Q_0),Q_0]}{\sqrt{n}}\cdot
e^{-n\sup_{s\ge 0}F(s,Q_0)}\nonumber\\
&=&(J-1)!\cdot\left(\frac{2\pi}{n}\right)^{(J-1)/2}\cdot\frac{1}{\sqrt{|\mbox{Hess}_F(Q_0)}|}
\frac{K_n[s_0(Q_0)),Q_0]}{\sqrt{n}}\cdot
e^{-n\min_Q\sup_{s\ge 0}F(s,Q)}\nonumber\\
&=&\frac{(J-1)!\cdot(2\pi)^{(J-1)/2}K_n[s_0(Q_0),Q_0]}{\sqrt{|\mbox{Hess}_F(Q_0)|}}
\cdot\frac{e^{-nR_d(D,\hat{P}_{\bx})}}{n^{J/2}}\nonumber\\
\end{eqnarray}
When $R_d(D,\hat{P}_{\bx})=0$, we have
\begin{eqnarray}
P_{\mbox{\tiny s}}^d[\bx]&\sim&\sum_{\{\hbx:~d(\bx,\hbx)\le nD\}}W(\hbx)\nonumber\\
&\sim&(J-1)!\cdot\int_{\{Q:~D_{\max}(Q)<D\}}\mbox{d}Q[1-o(n)]\nonumber\\
	&=&(J-1)!\cdot\mbox{Vol}\{Q:~D_{\max}(Q)<D\}\cdot[1-o(n)].
\end{eqnarray}
This completes the proof of Lemma \ref{lemma1}.

\section{Main Result}
\label{mainresult}

In the previous section, we focused on the evaluation of the probability that
a single randomly chosen codeword, under the mixture distribution, happens to
be successful in encoding a given source sequence, $\bx$, within distortion
$nD$. In this section, we harness the result of Lemma \ref{lemma1} for our
main coding theorem. The analysis, in this section, will be based on the
following simple well known fact: Let $\bx\in\calX^n$ be given and let $\hat{\bX}_1,\hat{\bX}_2,\ldots$ be a sequence of
$n$-vectors in $\hat{\calX}^n$, randomly and independently drawn under $W$.
Let $I_d(\bx)$
denote the index, $i$, of the first vector $\hat{\bX}_i$ with
$d(\bx,\hat{\bX}_i)\le nD$. Then, for every positive integer, $M$:
\begin{equation}
\label{fact1}
\mbox{Pr}\{I_d(\bx)> M\}=(1-P_{\mbox{\tiny
s}}^d[\bx])^M=\exp\{M\ln(1-P_{\mbox{\tiny s}}^d[\bx])\}\le \exp\{-M\cdot
P_{\mbox{\tiny s}}^d[\bx])\},
\end{equation}
and so, if $M=M_n= e^{\lambda_n}/P_{\mbox{\tiny s}}^d[\bx]$,
for some arbitrary positive sequence, $\{\lambda_n\}$, that tends to infinity,
then 
\begin{equation}
\label{fact2}
\mbox{Pr}\{I_d(\bx)> M_n\}\le \exp\{-e^{\lambda_n}\}.
\end{equation}
In particular, eq.\ (\ref{fact2}) holds if $R_d(D,\hat{P}_{\bx})> 0$ and
$M_n=\exp\{nR_d(D,\hat{P}_{\bx})+\frac{J\ln n}{2}+C+\lambda_n\}$, or if
$R_d(D,\hat{P}_{\bx})= 0$ and $M_n=\exp\{\lambda_n+C\}$, where $C>0$ is some
constant. We will make use of this fact several times in this section.

Consider next a randomly selected codebook of $A^n$
codewords, where $A$ in an arbitrary positive integer, strictly larger than
$\max\{J,K\}$, and where each codeword is drawn
independently under $W$.
Let the randomly selected codebook be revealed to both the encoder and the
decoder.

Consider next the following encoder.
Similarly as before, let $I_d(\bx)$ be defined as the
index of the first
codeword that falls within $d$-distortion $nD$ away from $\bx$, but now, with the small
twist that 
if none of the $A^n$ codewords fall within distortion $nD$ from $\bx$, then
we define $I_d(\bx)= L^n$ nevertheless (even though the distortion is larger
than $nD$).
Define the following probability distribution over the integers,
$1,2,\ldots,A^n$:
\begin{equation}
U[i]=\frac{1/i}{\sum_{k=1}^{A^n}1/k},~~~~i=1,2,\ldots,A^n.
\end{equation}
Given $\bx$ and distortion measure $d$, the encoder finds $I_d(\bx)$ and encodes it using a variable-rate
lossless code with the length function (in nats, and ignoring the equivalent
of the integer length
constraint),
\begin{eqnarray}
L_d(\bx)&=&-\ln U[I_d(\bx)]\nonumber\\
&\le&\ln I_d(\bx)+\ln\left(\sum_{k=1}^{A^n}\frac{1}{k}\right)\nonumber\\
&\le&\ln I_d(\bx)+\ln(\ln A^n+1)\nonumber\\
&=&\ln I_d(\bx)+\ln(n\ln A+1)\nonumber\\
&\le&\ln I_d(\bx)+\ln n +c,
\end{eqnarray}
where $c=\ln(\ln A+1)$.
Therefore, the expected codeword length for $\bx$ w.r.t.\ the randomness of the code
\begin{eqnarray}
\bE\{L_d(\bx)\}&\le&\bE\{\ln I_d(\bx)\}+\ln n+c\nonumber\\
&\le&\ln \bE\{I_d(\bx)\}+\ln n+c\nonumber\\
&=&\ln \left(\sum_{k=1}^{A^n}k\cdot\left(1-P_{\mbox{\tiny
s}}^d[\bx]\right)^{k-1}\cdot P_{\mbox{\tiny
s}}[\bx]+A^n\cdot\left(1-P_{\mbox{\tiny s}}[\bx]\right)^{A^n}\right)+\ln n+c\nonumber\\
&=&\ln \left(\sum_{k=1}^{\infty}\min\{k,A^n\}\cdot\left(1-P_{\mbox{\tiny
s}}^d[\bx]\right)^{k-1}\cdot P_{\mbox{\tiny
s}}^d[\bx]\right)+\ln n+c\nonumber\\
&\le&\ln\left\{\sum_{k=1}^\infty k\cdot\left(1-P_{\mbox{\tiny
s}}^d[\bx]\right)^{k-1}\cdot P_{\mbox{\tiny s}}^d[\bx]\right\}+\ln n+c\nonumber\\
&=&\ln\left(\frac{1}{P_{\mbox{\tiny s}}^d[\bx]}\right)+\ln n+c\nonumber\\
&\le&nR_d(D,\hat{P}_{\bx})+\left(\frac{J}{2}+1\right)\ln n+c',
\end{eqnarray}
where $c'$ is a constant, and where in the last step, we have used eq.\
(\ref{maincorollary}).

Our goal, in this section, however, is more ambitious than that. We wish to
prove the existence of a codebook with the following properties: 
(a) $L_d(\bx)$ is upper bounded
in terms of $R_d(D,\hat{P}_{\bx})$ plus some redundancy terms for every
$\bx\in\calX^n$ and bounded $d$, and (b)
The distortion
constraint is met
for every $\bx\in\calX^n$ and every distortion measure $d$ with a given $d_{\max}
<\infty$. 
To prove the second property, our approach is similar to that of Mahmood and
Wagner \cite{MW22a}:
We consider a fine grid, $\calD_n$, in the space of distortion matrices,
$\calD=[0,d_{\max}]^{JK}$, where for each entry of the distortion matrix, there
are $n$ grid points with spacings of $d_{\max}/n$, that is
$\calD_n=\{0\cdot d_{\max}/n,1\cdot d_{\max}/n,2\cdot d_{\max}/n,\ldots,n\cdot d_{\max}/n\}^{JK}$. If we can
prove that there exists a codebook where property (b) holds just for every $d\in\calD_n$, then for every
$d\in\calD$, the distortion cannot exceed $D+d_{\max}/n$. It should be pointed
out that the choice of $n$ as
the number of grid points for each entry $d$ is rather arbitrary, and can be
viewed just as an example. In fact, one
can afford even an exponentially fine resolution (and hence an exponentially
decaying distortion redundancy),
and our result will still hold. 
In spite of the similarity to Mahmood and Wagner's approach, there is an
important difference:
In our case, the quantization of the distortion measure takes part
only in the proof itself, not in the actual codebook construction, as in
\cite{MW22a}.

Our main coding theorem, in this work, is the following.
\begin{theorem}
\label{thm1}
Let $\epsilon > 0$ be arbitrarily small. For all sufficiently large $n$, there
exists a codebook $\calC_n=\{\hbx_1,\hbx_2,\ldots,\hbx_{A^n}\}$, such that for
every $\bx\in\calX^n$ and every $d\in\calD_n$, the following two properties
hold at the same time:
\begin{enumerate}
\item[(a)] If $R_d(D,\hat{P}_{\bx})> 0$,
\begin{equation}
\frac{L_d(\bx)}{n}\le
R_d(D,\hat{P}_{\bx})+\left(\frac{J}{2}+2+\epsilon\right)\cdot\frac{\ln
n}{n}+o\left(\frac{\ln n}{n}\right).
\end{equation}
If $R_d(D,\hat{P}_{\bx})=0$,
\begin{equation}
\frac{L_d(\bx)}{n}\le
(2+\epsilon)\cdot\frac{\ln
n}{n}+o\left(\frac{\ln n}{n}\right).
\end{equation}
\item[(b)] $d(\bx,\hbx)\le nD$.
\end{enumerate}
\end{theorem}
The main redundancy term in part (a), namely,
$$\left(\frac{J}{2}+2+\epsilon\right)\frac{\ln
n}{n},$$
should be compared with those of Mahmood and Wagner \cite{MW22a}, where the coefficients
in front of $(\ln n)/n$ are, respectively, $2JK+J+3$,
$JK+J$, and
$J^2K^2+J-2$, in Theorems 1, 2, and 3 of \cite{MW22a}. The differences are
quite significant, especially for large $J$ and $K$.

The remaining part of this section is devoted to the proof of Theorem
\ref{thm1}.

\noindent
{\em Proof of Theorem \ref{thm1}.}
In this proof, we confine attention only to the more interesting case where $R_d(D,\hat{P}_{\bx})>
0$, but the case $R_d(D,\hat{P}_{\bx})=0$ can easily be handled in the very same
manner. Consider the quantity
\begin{eqnarray}
E_n&\dfn&\bE\bigg\{\max\bigg(\max_{d\in\calD_n}\max_{\bx\in\calX^n}\calI\{d(\bx,\hat{\bX})>
nD\},\nonumber\\
& &\left[\max_{d\in\calD_n}\max_{\bx\in\calX^n}\left(L_d(\bx)-
nR_d(D;\hat{P}_{\bx})-\left(\frac{J}{2}+2+\epsilon\right)\ln
n-c\right)\right]_+\bigg)\bigg\},
\end{eqnarray}
where the expectation is w.r.t.\ the randomness of the code, $\calC_n$.
If we can bound $E_n$ by a sequence, $\delta_n$, that decays
as $n\to\infty$, this will imply that there must exist a
code for which both
\begin{equation}
\label{dist}
\max_{d\in\calD_n}\max_{\bx\in\calX^n}\calI\{d(\bx,\hbx)>
nD\}\le\delta_n
\end{equation}
and
\begin{equation}
\label{length}
\max_{d\in\calD_n}\max_{\bx\in\calX^n}\left(L_d(\bx)-
nR_d(D;\hat{P}_{\bx})-\left(\frac{J}{2}+2+\epsilon\right)\log
n-c\right)\le\delta_n
\end{equation}
at the same time. 
Observe that since the left-hand side of (\ref{dist}) is either zero or one,
then if we know that 
it must be less than $\delta_n\to 0$, for some codebook, $\calC_n$, it means
that it must vanish
as soon as $n$ is large enough such that $\delta_n < 1$, namely $d(\bx,\hbx)\le nD$ for all $\bx\in\calX^n$ and
$d\in\calD_n$. Also, by (\ref{length}), for the same codebook, we must have
\begin{equation}
L_d(\bx)\le nR_d(D;\hat{P}_{\bx})+\left(\frac{J}{2}+2+\epsilon\right)\ln
n+c+\delta_n~~~~\forall~\bx\in\calX^n,~d\in\calD_n,
\end{equation}
where the extra term, $\delta_n$, adds a negligible amount to the redundancy.

To prove that $E_n$ decays, we begin with the simple fact
that the maximum between two non-negative numbers is upper bounded by their
sum, which implies that
\begin{eqnarray}
\label{sum}
E_n&\le&\bE\left\{\max_{d\in\calD_n}\max_{\bx\in\calX^n}\calI\{d(\bx,\hat{\bX})>
nD\}\right\}+\nonumber\\
& &\bE\left\{\left[\max_{d\in\calD_n}\max_{\bx\in\calX^n}\left(L_d(\bx)-
nR_d(D,\hat{P}_{\bx})-\left(\frac{J}{2}+2+\epsilon\right)\ln
n-c\right)\right]_+\right\},
\end{eqnarray}
and so, it is enough to prove that each one of the terms decays with $n$.
As for the first term, we have:
\begin{eqnarray}
& &\bE\left\{\max_{d\in\calD_n}\max_{\bx\in\calX^n}\calI\{d(\bx,\hat{\bX})>
nD\}\right\}\nonumber\\&\le&
\bE\left\{\sum_{d\in\calD_n}\sum_{\bx\in\calX^n}\calI\{d(\bx,\hat{\bX})>
nD\}\right\}\nonumber\\
&=&\sum_{d\in\calD_n}\sum_{\bx\in\calX^n}\bE\left\{\calI\{d(\bx,\hat{\bX})>
nD\}\right\}\nonumber\\
&=&\sum_{d\in\calD_n}\sum_{\bx\in\calX^n}\mbox{Pr}\{d(\bx,\hat{\bX})>
nD\}\nonumber\\
&=&\sum_{d\in\calD_n}\sum_{\bx\in\calX^n}\left(1-P_{\mbox{\tiny
s}}^d[\bx]\right)^{A^n}\nonumber\\
&\le&\sum_{d\in\calD_n}\sum_{\bx\in\calX^n}\exp\left\{-A^nP_{\mbox{\tiny
s}}^d[\bx]\right\}\nonumber\\
&\le&\sum_{d\in\calD_n}\sum_{\bx\in\calX^n}\exp\left\{-\exp\left\{n\left[\ln
A-R_d(D,\hat{P}_{\bx})-O\left(\frac{\ln
n}{n}\right)\right]\right\}\right\}\nonumber\\
&\le&(n+1)^{JK}\cdot J^n\exp\left(-\exp\left\{n\left[\ln A-\ln
J-O\left(\frac{\ln n}{n}\right)\right]\right\}\right),
\end{eqnarray}
which indeed decays as $n\to\infty$, since we have assumed that $A > J$.
As for the second term of (\ref{sum}), we have:
\begin{eqnarray}
& &\bE\left\{\left[\max_{d\in\calD_n}\max_{\bx\in\calX^n}\left(L_d(\bx)-
nR_d(D,\hat{P}_{\bx})-\left(\frac{J}{2}+2+\epsilon\right)\ln
n-c\right)\right]_+\right\}\nonumber\\
&=&\bE\left\{\left[\max_{d\in\calD_n}\max_{\bx\in\calX^n}\left(\ln I_d(\bx)
-nR_d(D,\hat{P}_{\bx})-\left(\frac{J}{2}+1+\epsilon\right)\ln
n\right)\right]_+\right\}\nonumber\\
&=&\int_0^\infty\mbox{Pr}\left\{\max_{d\in\calD_n}\max_{\bx\in\calX^n}\left[\ln
I_d(\bx)-
nR_d(D,\hat{P}_{\bx})-\left(\frac{J}{2}+1+\epsilon\right)\ln
n\right]\ge s\right\}\mbox{d}s\nonumber\\
&=&\int_0^{n\ln A}
\mbox{Pr}\left\{\max_{d\in\calD_n}\max_{\bx\in\calX^n}\left[\ln
I_d(\bx)-
nR_d(D,\hat{P}_{\bx})-\left(\frac{J}{2}+1+\epsilon\right)\ln
n\right]\ge s\right\}\mbox{d}s\nonumber\\
&\le&\int_0^{n\ln A}
\mbox{Pr}\bigcup_{d\in\calD_n}\bigcup_{\bx\in\calX^n}\left\{I_d(\bx)\ge
\exp\left[nR_d(D,\hat{P}_{\bx})+\left(\frac{J}{2}+1+\epsilon\right)\ln
n+s\right]\right\}\mbox{d}s\nonumber\\
&\le&\sum_{d\in\calD_n}\sum_{\bx\in\calX^n}\int_0^{n\ln A}
\mbox{Pr}\left\{I_d(\bx)\ge
\exp\left[nR_d(D,\hat{P}_{\bx})+\left(\frac{J}{2}+1+\epsilon\right)\ln
n+s\right]\right\}\mbox{d}s\nonumber\\
&\le&\sum_{d\in\calD_n}\sum_{\bx\in\calX^n}\int_0^{n\ln A}
\mbox{Pr}\left\{I_d(\bx)\ge
\exp\left[nR_d(D,\hat{P}_{\bx})+\left(\frac{J}{2}+1+\epsilon\right)\ln
n\right]\right\}\mbox{d}s\nonumber\\
&\le&(n\ln A)\cdot\sum_{d\in\calD_n}\sum_{\bx\in\calX^n}(1-P_{\mbox{\tiny s}}^d[\bx])^{\exp
\left[nR_d(D,\hat{P}_{\bx})+\left(\frac{J}{2}+1+\epsilon\right)\ln
n\right]}\nonumber\\
&=&(n\ln A)\cdot\sum_{d\in\calD_n}\sum_{\bx\in\calX^n}
\exp\left\{\exp
\left[nR_d(D,\hat{P}_{\bx})+\left(\frac{J}{2}+1+\epsilon\right)\ln
n\right]\ln(1-P_{\mbox{\tiny s}}^d[\bx])\right\}\nonumber\\
&\le&(n\ln A)\cdot\sum_{d\in\calD_n}\sum_{\bx\in\calX^n}\exp\left\{-\exp
\left[nR_d(D,\hat{P}_{\bx})+\left(\frac{J}{2}+1+\epsilon\right)\ln
n\right]P_{\mbox{\tiny s}}^d[\bx]\right\}\nonumber\\
&\le&(n\ln
A)\cdot\sum_{d\in\calD_n}\sum_{\bx\in\calX^n}\exp\left\{-\exp\left[(1+\epsilon)\ln
n\right]\right\}\nonumber\\
&=&(n\ln A)\cdot(n+1)^{JK}\cdot J^n\cdot\exp\{-n^{1+\epsilon}\},
\end{eqnarray}
which decays as well.
This completes the proof of Theorem \ref{thm1}.

\section{Beyond Memoryless Sources and Additive
Distortion Measures}
\label{LZ}

Our results in Sections \ref{mainlemma} and \ref{mainresult} hold {\em
pointwise}, for each and every individual source vector $\bx$, even without taking the expectation w.r.t.\
the randomness of the source vector. Of course, one can also take the
expectation and obtain a result on the rate redundancy relative to the expectation
of the empirical rate-distortion function, $R_d(D,\hat{P}_{\bx})$ (which
in turn converges almost surely to $R_d(D,P)$), as was actually done in \cite[Theorems
1--4]{MW22b}. But in spite of the pointwise nature of our results so far, the
codes that we have been considering are suitable only for the class of  memoryless sources and
additive distortion measures, since
the length function, $L_d(\bx)$, whose main term is $nR_d(D,\hat{P}_{\bx})$,
depends on $\bx$ only via its zeroth order empirical distribution, which is
blind to any empirical dependencies and repetitive patterns within the source
sequence, $\bx$. 

In this section, we would like to remain in the realm of individual sequences,
but to expand the scope to codes that are suitable beyond memoryless
sources, i.e., codes that are designed to exploit the memory within
the given source sequence to be compressed. By the same token, we will be
interested in more general classes of distortion measures, not necessarily
additive ones. In this section, the discussion will be less formal than
before, as we will only outline how the ideas of the
previous sections extend to this more general setting, without any heavy
analysis of exact redundancy rates.

We adopt the individual-sequence setting, in the
footsteps of Lempel
and Ziv \cite{ZL78}. According to this setting, defined in \cite{ZL78}
for the lossless case, the source sequence,
$\bx$, is a given deterministic setting, but the encoder is limited to be
implementable by an information lossless finite-state machine with $s$ states, and the asymptotic
regime is that $s \ll n$, as the limit $s\to\infty$ is taken after the limit
$n\to \infty$. 

When it comes to source coding with distortion, a natural extension of this
setting could be based on the fact that in lossy compression, there is no loss
of optimality if the encoder is implemented as a cascade of two mappings, as follows: first, apply a reproduction
encoder (or, vector quantizer), that maps the source $\bx$ directly to its
reproduction, $\hbx$, and then compress $\hbx$ by a lossless encoder, without
any additional distortion. Accordingly, we can adopt this structure with the
limitation that the lossless encoder of the second stage is a finite-state
encoder with $s$ states, exactly as in \cite{ZL78}.\footnote{Note that this setting is
somewhat different form Ziv's model of lossy compression for individual
sequences, \cite{Ziv80}.} Applying, the converse theorem of Lempel and Ziv
\cite[Theorem 1]{ZL78}, we have that the length of the lossless code associated with
the reproduction vector, $\hbx$, is lower bounded by
\begin{equation}
L(\hbx)\ge [c(\hbx)+s^2]\log\frac{c(\hbx)+s^2}{4s^2}+2s^2,
\end{equation}
where $c(\hbx)$ is the largest number of distinct phrases whose concatenation
forms $\hbx$. Since $\hbx$ is constrained to lie within distance $nD$ away
from $\bx$, we reach at the obvious lower bound of
\begin{equation}
\label{lzlowerbound}
L_d(\bx)\ge \min_{\{\hbx:~d(\bx,\hbx)\le
nD\}}\left\{[c(\hbx)+s^2]\log\frac{c(\hbx)+s^2}{4s^2}+2s^2\right\},
\end{equation}
and a conceptually simple way to asymptotically achieve this lower bound is to
choose, among all vectors, $\{\hbx\}$, within distortion $nD$ away from $\bx$,
the one whose Lempel-Ziv (LZ) code-length is minimal, and to transmit its
compressed from using the LZ algorithm \cite[Theorem 2]{ZL78}. The LZ
codelength of $\hbx$, which we denote by $LZ(\hbx)$, is upper bounded by
\begin{equation}
LZ(\hbx)\le [c_{\mbox{\tiny LZ}}(\hbx)+1]\log(2K[c_{\mbox{\tiny
LZ}}(\hbx)+1]),
\end{equation}
where $c_{\mbox{\tiny LZ}}(\hbx)$ is the number of phrases of $\hbx$ obtained
by the incremental parsing procedure of the LZ algorithm \cite[proof of Theorem
2]{ZL78}.
Note that here, $d(\bx,\hbx)$ can be any distortion function, not
necessarily an additive one.
The painful
part of this achievability scheme, however, is the exponential
complexity associated with the search across the `sphere',
$\{\hbx:~d(\bx,\hbx)\le nD\}$. In the case of an additive distortion measure,
the complexity of this search grows at the exponential rate
of $\exp\{nE(D)\}$, where $E(D)=\max H(\hat{X}|X)$, with $X$ being a dummy random
variable, distributed according to $\hat{P}_{\bx}$, and with the maximization
being taken over all conditional distributions, $\{P_{\hat{X}|X}\}$, such that
$\bE\{d(X,\hat{X})\}\le D$. When $D$ is relatively large, then so is $E(D)$.

We now propose an alternative approach to this problem using the ideas of the previous section. To this
end, we first have to extend the random coding distribution, $W$, to be
suitable beyond the class of memoryless sources. Following the findings of
\cite{CM21} and \cite{MC20}, consider the random coding distribution,
\begin{equation}
W(\hbx)=\frac{2^{-LZ(\hbx)}}{\sum_{\hbx'\in\hat{\calX}^n}2^{-LZ(\hbx')}}.
\end{equation}
The associated single success probability is given by
\begin{equation}
\label{LZsuccess}
P_{\mbox{\tiny s}}^d[\bx]=\sum_{\{\hbx:~d(\bx,\hbx)\le nD\}}
W(\hbx).
\end{equation}
We can repeat the same derivations as in Section \ref{mainresult}, but with the new expression of
$P_{\mbox{\tiny s}}^d[\bx]$, and use eqs.\ (\ref{fact1}) and (\ref{fact2}) to
argue that we can achieve compression according to the length function, 
\begin{equation}
L_d(\bx)=-\log P_{\mbox{\tiny
s}}^d[\bx]+(2+\epsilon)\log n 
\end{equation}
within distortion $nD$ (w.r.t.\ any distortion measure $d$ within a class
$\calD$ of distortion measures that can be well approximated using a grid
whose size is no more than exponential), pointwise, for every
$\bx$, similarly as before. Now, observe that
the main term of $L_d(\bx)$, namely, $-\log P_{\mbox{\tiny s}}^d[\bx]$, can be
upper bounded as follows.
\begin{eqnarray}
-\log P_{\mbox{\tiny s}}^d[\bx]&=&-\log\left[\sum_{\{\hbx:~d(\bx,\hbx)\le
nD\}}\frac{2^{-LZ(\hbx)}}{\sum_{\hbx'}2^{-LZ(\hbx')}}\right]\nonumber\\
&\le&-\log\left[\sum_{\{\hbx:~d(\bx,\hbx)\le
nD\}}2^{-LZ(\hbx)}\right]\nonumber\\
&\le&-\log\left[\max_{\{\hbx:~d(\bx,\hbx)\le
nD\}}2^{-LZ(\hbx)}\right]\nonumber\\
&=&\min_{\{\hbx:~d(\bx,\hbx)\le nD\}}LZ(\hbx),
\end{eqnarray}
where in the second line, we used Kraft's inequality.
This means that this scheme also asymptotically achieves the lower bound
(\ref{lzlowerbound}). However, this coding scheme has a different computational complexity
than the earlier one. The number of metric calculations that this encoder has
to carry out until it finds the first codeword within distortion $nD$, is a random variable, but it is typically of the order of
magnitude of $1/P_{\mbox{\tiny s}}^d[\bx]$.
Which one of the encoders is better in terms of the computational complexity?
The answer depends, of course, on $\bx$ and $D$. For small $D$, it is more efficient to use the
first approach, as $e^{nE(D)}$
is relatively small, whereas $1/P_{\mbox{\tiny s}}^d[\bx]$ is
relatively large. On the other hand, for large $D$, the contrary is true. 
In fact, by Ziv's inequality \cite[p.\ 455, eq.\ (13.125)]{CT06} (applied to
memoryless sources), it is readily seen that $1/P_{\mbox{\tiny s}}^d[\bx]\lexe \exp\{nR(D,\hat{P}_{\bx})\}$,
and so, whenever $R(D,\hat{P}_{\bx})< E(D)$,
it is definitely better to use the second scheme.

\end{document}